\documentclass[12pt]{iopart}

\usepackage{iopams}

\expandafter\let\csname equation*\endcsname\relax

\expandafter\let\csname endequation*\endcsname\relax

\usepackage{amsmath}

\usepackage[usenames]{color}

\newcommand{\be}{\begin{equation}}
\newcommand{\ee}{\end{equation}}
\newcommand{\bea}{\begin{eqnarray}}
\newcommand{\eea}{\end{eqnarray}}

\definecolor{bc}{rgb}{0, 0.7, 0.0}

\newcommand{\ud}{\mathrm{d}}
\newcommand{\LCm}{{\scriptscriptstyle -}} 
\newcommand{\LCp}{{\scriptscriptstyle +}}
\newcommand{\LCpm}{{\scriptscriptstyle \pm}}

\newcommand{\LCperp}{{\scriptscriptstyle \perp}}

\begin{document}

\title[Superintegrable relativistic systems]{Superintegrable relativistic systems in spacetime-dependent background fields}

\author{T.~Heinzl and A.~Ilderton}
\address{Centre for Mathematical Sciences, University of Plymouth, PL4 8AA, UK}
\eads{\mailto{thomas.heinzl@plymouth.ac.uk}, \mailto{anton.ilderton@plymouth.ac.uk}}
\vspace{10pt}

\begin{abstract}
We consider a relativistic charged particle in background electromagnetic fields depending on both space and time. We identify which symmetries of the fields automatically generate integrals (conserved quantities) of the charge motion, accounting fully for relativistic and gauge invariance. Using this we present new examples of superintegrable relativistic systems. This includes examples where the integrals of motion are quadratic or nonpolynomial in the canonical momenta.
\end{abstract}

\vspace{2pc}

\section{Introduction}
The majority of known superintegrable systems correspond to dynamics on the low-dimensional Euclidean spaces $E_2$ or $E_3$ and are non-relativistic, see~\cite{Miller:2013} for a recent review.  Here we present new examples of superintegrable systems in the relativistic dynamics of charged particles in background electromagnetic fields with nontrivial space-time dependendence.

Recall that a classical system with $2n$-dimensional phase space is integrable if it admits $n$ conserved quantities $Q_j$ which are functionally independent and in involution, so $\{Q_i,Q_j\}=0\, \forall\, i,j = 1, \ldots , n$, see e.g.\ \cite{Kozlov:1983,Dubrovin:1990}. For autonomous systems the Hamiltonian itself, $H$, may be taken as one of the $Q_j$. If there are a further $k$ conserved quantities, $1\leq k\leq n-1$, then the system is superintegrable~\cite{Wojciechowski:1983}. If $k=1$ the system is minimally superintegrable, if $k=n-1$ it is maximally superintegrable.

These definitions may seem to present an obstacle in the case of relativistic systems, as reparameterisation invariance of the relativistic particle action implies that the Hamiltonian is identically zero~\cite{Dirac:1950pj}. The solution to this problem is though well known; one `gauge fixes' the reparameterisation invariance and singles out a preferred time co-ordinate~\cite{Heinzl:2000ht}. The disadvantage (from a physicist's perspective) is that in doing so one loses manifest Lorentz invariance, but the benefit is that by choosing a preferred time one obtains a well-defined Hamiltonian system.

Although the purpose of this paper is to flag the existence of novel relativistic superintegrable examples, expanding the literature, the methods behind these examples could be turned into a systematic study exhausting all possibilities, following e.g.~\cite{vanHolten:2006xq,Marchesiello:2015}.

This paper is organised as follows. We begin in Sect.~\ref{SECT:REVIEW} by briefly reviewing the necessary elements of relativistic particle dynamics in the Lagrangian and Hamiltonian formalisms. We show here that if a background field is symmetric under a Poincar\'e transformation then charge motion in that field automatically admits a related conserved quantity. Based on this result we present a number of superintegrable relativistic systems in Sections~\ref{SECT:SUPER1} and \ref{SECT:SUPER2}, in order of decreasing number of Poincar\'e symmetries. We conclude in Sect~\ref{SECT:CONCS}, where we also comment on the extension of our results to quantum mechanics.

%
\section{Relativistic dynamics}\label{SECT:REVIEW}
%
\subsection{Integrals of motion from Poincar\'e symmetries}
We consider a relativistic particle of unit mass and charge with spacetime coordinates~$x^\mu(\tau)$ moving in a background electromagnetic field $A_\mu(x)$. The relativistic particle action is
\be \label{ACTION}
	S = \int\!\ud \tau \; L = -\int\!\ud \tau \; \big( \sqrt{{\dot x}^\mu {\dot x}_\mu} +  \dot{x}^\mu A_\mu(x)\big) \;,
\ee
where $\tau$ is proper time and $\dot{x}^\mu\equiv \ud x^\mu /\ud \tau$. Varying the action functional one finds that it becomes stationary for particle worldlines obeying the Lorentz equation of motion,
\be \label{LORENTZ}
  \dot{p}_\mu = \dot{x}^\nu \partial_\mu A_\nu \; , 
\ee
in which $p_\mu$ is the canonical momentum defined by
\be
	p_\mu = -\frac{\partial L}{\partial \dot{x}^\mu} = \frac{\dot{x}_\mu}{\sqrt{\dot{x}^2}} + A_\mu(x) \;.
\ee
(The minus sign ensures the correct nonrelativistic limit in our conventions.)  A free particle, $A_\mu=0$, has Poincare\'e symmetry. The infinitesimal form of a Poincar\'e transformation is described by $\xi_\mu(x) = a_\mu + \omega_{\mu\nu}x^\nu$, where $a_\mu$ and $\omega_{\mu\nu}$ are constant, and $\omega_{\mu\nu} = - \omega_{\nu\mu}$;  these parametrise, respectively, the four translations and six Lorentz transformations comprising the Poincar\'e group (and corresponding to the 10 Killing vectors of flat Minkowski space~\cite{Stephani:2004ud}). In the free theory, the Poincar\'e symmetry is generated by 10 conserved `Noether charges' $\xi.p$. In the presence of the background $A_\mu$, however, these acquire a proper time dependence, which may be found directly from the equation of motion (\ref{LORENTZ}), 
\be\label{invarians}
	\frac{\ud}{\ud \tau} \xi.p = \dot{x}^\mu\mathcal{L}_\xi A_\mu \;,
\ee
where $\mathcal{L}_\xi$ is the Lie derivative of the background field under the Poincar\'e transform,
\be\label{delta-A}
	\mathcal{L}_\xi A_\mu  \equiv \xi.\partial A_\mu + A_\nu \partial_\mu \xi^\nu \;.
	\ee
$A_\mu$ is called `symmetric' if it is invariant under the action of the Lie derivative up to a $U(1)$ gauge transformation~\cite{Jackiw:1995be}, i.e.~if
\be
	\mathcal{L}_\xi A_\mu = \partial_\mu \Lambda \;,
\ee
where the scalar field $\Lambda$ depends implicity on $\xi$. This is equivalent to the physical electromagnetic fields, $F_{\mu\nu}=\partial_\mu A_\nu - \partial_\nu A_\mu$, being strictly invariant,
\be
	\mathcal{L}_\xi F_{\mu\nu} \equiv  \xi.\partial F_{\mu\nu} + F_{\sigma\nu} \partial_\mu \xi^\sigma +  F_{\mu\sigma} \partial_\nu \xi^\sigma= 0 \;.
\ee
If $A_\mu$ is symmetric then (\ref{invarians}) becomes an exact differential with respect to $\tau$ and can be integrated. Thus a Poincar\'e symmetric background automatically implies an integral of motion (conserved quantity)~$Q$,
\be\label{Q-GAUGE}
	Q = \xi.p - \Lambda \;.
\ee
It follows that if we can identify a background with sufficiently many Poincar\'e symmetries, charge dynamics in that background will be (super) integrable. To make this concrete, we turn to the Hamiltonian picture.

\subsection{Hamiltonian formulation of relativistic dynamics}
The action (\ref{ACTION}) is the proper time integral of a Lagrangian which is homogeneous of first degree in velocities, $L[\lambda \dot{x}] = \lambda L[\dot{x}]$. Euler's homogeneous function theorem then implies that the Hamiltonian vanishes~\cite{Dirac:1950pj} (see~\cite{Salisbury:2016} for historical context and additional references), as is easily verified:  
\be \label{H=0}
	H = -p_\mu \dot{x}^\mu - L = 0 \;.
\ee
(Again, minus signs follows from conventions.) This has long been understood to be due to the reparametrisation invariance of~(\ref{ACTION}) under $\tau \to f(\tau)$, $\dot{f} > 0$. The most convenient solution to this problem is to give up manifest Lorentz covariance and `gauge fix' the reparametrisation invariance by choosing $\tau$ to be a physical time coordinate.  (This is unrelated to the gauge choice for the background  potential $A_\mu$.)  The details are not relevant here, only that there are basically three choices of time (as pointed out by Dirac \cite{Dirac:1949cp}), leading to different possible Hamiltonians, all of which should give equivalent descriptions of the dynamics. The choice is dictated by the symmetries of the system under consideration.  To gauge fix, one chooses a time variable, $\tau = G(x)$ and identifies the Hamiltonian as the variable conjugate to $G(x)$\footnote{The Poisson bracket between $H$ and $G$ is a Faddeev-Popov expression which must be nonzero to avoid gauge fixing ambiguities (`Gribov problems'~\cite{Gribov:1977wm}) which would render the Hamiltonian flow, hence time evolution, ill-defined, cf.~\cite{Heinzl:2000ht,Amaral:2016hud}. This could possibly also spoil the equivalence of different gauge choices.}. The two gauge fixings we will find useful in this paper are given below.
\begin{enumerate}
	\item \underline{Instant form} \\
	We gauge fix $\tau = t$, i.e.~time is $t$. Phase space is then six dimensional, spanned by the spatial co-ordinates, $x^j=(x,y,z)$, and their conjugate momenta, $p_j=(p_1,p_2,p_3)$.  The Poisson bracket is
\be
	\{X,Y\} = \frac{\partial X}{\partial x^j}\frac{\partial Y}{\partial p_j} -\frac{\partial X}{\partial p_j}\frac{\partial Y}{\partial x^j}  \;.
\ee
The Hamiltonian is the generator of evolution in $t$, namely~$p_0$, 
\be
	H = p_0 = \sqrt{1+(p_j - A_j)^2} + A_0 \;,
\ee
which can be simplified by adopting Weyl gauge, $A_0=0$. The time evolution of any quantity $Q$ is determined by
\be
	\frac{\ud Q}{\ud t} = \frac{\partial Q}{\partial t} - \{Q,H\} \;,
\ee
where we have allowed for explicit time dependence since the Hamiltonian will typically depend explicitly on time, through the background field.

The advantage of the instant form is that the time and canonical phase space variables are those familiar from non-relativistic mechanics. The disadvantage is the complicated square root in the Hamiltonian. In fact the majority of the superintegrable systems we will present are better discussed using the `front form'~\cite{Heinzl:2000ht,Neville:1971uc,Bakker:2013cea}, to which we now turn.  
\item \underline{Front form} \\
We gauge fix $\tau = x^\LCp\equiv t+z$, i.e.~time is $x^\LCp$. Phase space is six dimensional, spanned by the `longitudinal' coordinate $x^\LCm\equiv t-z$, `transverse' coordinates $x^\LCperp\equiv (x,y)$, and their conjugate momenta $p_\LCm$ and $p_\LCperp\equiv(p_1,p_2)$. The Poisson bracket is
\be
	\{A,B\} = 
	\frac{\partial A}{\partial x^\LCm}\frac{\partial B}{\partial p_\LCm} 
	 -
	 \frac{\partial A}{\partial p_\LCm} \frac{\partial B}{\partial x^\LCm}
	+
	 \frac{\partial A}{\partial x^\LCperp}\frac{\partial B}{\partial p_\LCperp} 
	 -
	  \frac{\partial A}{\partial p_\LCperp} \frac{\partial B}{\partial x^\LCperp}  \;.
\ee
The Hamiltonian is the generator of evolution in $x^\LCp$, namely $p_\LCp$, 
\be\label{H-LF}
	H= p_\LCp = \frac{(p_\LCperp - A_\LCperp(x))^2+1}{4(p_\LCm - A_\LCm)} + A_\LCp(x) \;.
\ee
Choosing light-front gauge, $A_\LCm=0$, simplifies the denominator. Time evolution is determined by
\be\label{utveckling-lf}
	\frac{\ud Q}{\ud x^\LCp} = \frac{\partial Q}{\partial x^\LCp} - \{Q,H\} \;.
\ee
We list below a convenient basis of the 10 Poincar\'e generators $P_\mu$ and $M^{\mu\nu} = x^\mu P^\nu - x^\nu P^\mu$ in the canonical variables of the front form, in which we will mostly work.
\bea
  \text{4 translations ($P_\mu$): }\qquad\quad &p_\LCp &\equiv H \, , \; p_\LCm \, , \; p_\LCperp \, ,   \\
  \text{1 rotation ($M^{12}$): } \quad &L_z &\equiv x p_2 - y p_1 \; ,
  \\  
  \text{1 boost ($M^{\LCm\LCp}$): } &K_z &\equiv x^\LCp p_\LCm - x^\LCm H  \, ,
  \\
   \text{2 null rotations ($M^{i\LCp}$): } &T_i &\equiv  2x^i p_\LCm + x^\LCp p_i \; ,  \label{T-DEF}
   \\
  \text{2 null rotations ($M^{i\LCm}$): } &U_i &\equiv  2x^i H + x^\LCm p_i \; .
\eea
\end{enumerate}

\section{Superintegrable systems with more than three Poincar\'e symmetries}\label{SECT:SUPER1}

\subsection{Plane waves}
We work in the front form. The gauge potential for a plane wave can be taken to have only two nonzero components,
\be\label{A-PW}
	A_j(x) = f'_j(x^\LCp) \;, \quad j \in\{1,2\} \;,
\ee
in which the $f_j$ are arbitrary functions and the prime, an $x^\LCp$-derivative, is for notational convenience. The Hamiltonian $H=p_\LCp$, (\ref{H-LF}), is then explicitly time ($x^\LCp$) dependent. The translational invariance of (\ref{A-PW}) leads to the conservation of all three of the canonical momenta:
\be
	Q_1 = p_1 \;, \quad Q_2 = p_2 \;, \quad Q_3 = p_\LCm \;,  
\ee
which are in involution, i.e.~the system is integrable~\cite{Landau:1987}.

What has seemingly not been noticed before is that a plane wave is invariant under the action of the two null rotations (\ref{T-DEF}), i.e.~$\mathcal{L}_\xi F_{\mu\nu}=0$, for $\xi$ such that $\xi.p = T_1$ or $T_2$. This implies that there are two further integrals of motion. It is easily verified that the potential (\ref{A-PW}) is, under $T_j$, symmetric up to a gauge transformation with $\mathcal{L}_{\xi} A_\mu = \partial_\mu f_j(x^\LCp)$. Hence the two additional integrals $Q_4$ and $Q_5$ are, combining (\ref{T-DEF}) and (\ref{Q-GAUGE})
\be\label{Q45}
	Q_4 = 2x p_\LCm + x^\LCp p_1 - f_1(x^\LCp)  \;, \qquad
	Q_5 = 2y p_\LCm + x^\LCp p_2 - f_2(x^\LCp) \;.
\ee
This may be verified directly by taking Poisson brackets with $H$. Thus a particle in a background plane wave is a \textit{maximally superintegrable} relativistic system. $Q_4$ and $Q_5$ are in involution with each other and with $Q_1$, but not with $Q_2$ and $Q_3$.

The solution of the equations of motion proceeds as follows. All three momenta are conserved. From the conservation of $Q_4$ and $Q_5$ in (\ref{Q45}) we are able to read off the transverse orbits immediately:
\be
	x(x^\LCp) = \frac{Q_4 + f_1(x^\LCp) - Q_1 x^\LCp}{2 Q_3} \;,  \qquad y(x^\LCp) = \frac{Q_5 + f_2(x^\LCp) - Q_2 x^\LCp}{2 Q_3} \;.
\ee
It remains only to identify $x^\LCm$, Hamilton's equation for which is
\be\label{decoupled-PW}
	\frac{\ud x^\LCm}{\ud x^\LCp} = -\{x^\LCm,H\} = \frac{1+\big(Q_1 - f_1'(x^\LCp)\big)^2+ \big(Q_2 - f_2'(x^\LCp)\big)^2}{4Q^2_3} \;.
\ee
This can be integrated directly. The orbits given by this elegant method agree exactly with those found by standard methods, see~\cite{Heinzl:2008rh} and references therein.

\subsection{TM-mode model}
For our next example we consider fields which are symmetric under translations in $x^\LCm$ and under two null rotations, as for plane waves, but we abandon transverse translation invariance.  A potential symmetric (without gauge term) under $p_\LCm$ and $T_j$ is 
\be\label{A-axicon}
	A_\LCp = -\frac{x^\LCperp x^\LCperp}{2x^{\LCp 2}}f(x^\LCp) \;, \quad
	A_\LCm = -\frac{1}{2}f(x^\LCp) \;, \quad
	A_\LCperp = \frac{x^\LCperp}{x^\LCp} f(x^\LCp) \;,
\ee
where $f$ is arbitrary. The electric and magnetic fields are, for $\mathcal{E}(x) := f'(x^\LCp)/x^\LCp$,
\be
	{\bf E} = \mathcal{E}(x^\LCp) \big( x , y , x^\LCp \big)  \;, \qquad {\bf B} = \mathcal{E}(x^\LCp) \big( y , -x , 0 \big) \;.
\ee
The fields describe a radially (azimuthally) polarised electric (magnetic) field transverse to the propagation direction, along with a longitudinal electric field. This is a toy model of a transverse magnetic (TM) laser beam near the beam axis~\cite{McDonald:2000b}.

As the potential (\ref{A-axicon}) is symmetric without gauge term, the three quantities
\be\label{Q123}
	Q_1 = 2x p_\LCm + x^\LCp p_1 \;, \qquad Q_2 = 2y p_\LCm + x^\LCp p_2 \;,  \qquad Q_3 = p_\LCm \;, 
\ee
are conserved under the action of the lightfront Hamiltonian as in (\ref{utveckling-lf}), and are in involution. Hence charge motion is integrable. One can verify directly that angular momentum
\be\label{Q4}
	Q_4 = L_z = x p_2 - y p_1 \;,
\ee
is also conserved, and so that charge motion becomes {\it minimally superintegrable} due to Poincar\'e symmetries. A direct computation shows that this exhausts the possible Poincar\'e symmetries $\xi$ of the field giving $\mathcal{L}_\xi F_{\mu\nu} =0$. However, there is a fifth integral, which is hence not related to Poincar\'e invariance. This is found, following~\cite{Miller:2013,Fris:1965,Marquette:2008}, by making the ansatz that $Q$ is of a certain order in the momenta, and then imposing (\ref{utveckling-lf}). Here, we begin by assuming that $Q$ is {\it linear} in $p_\LCperp$ but otherwise depends arbitrarily on the coordinates and on $p_\LCm$, i.e.~we assume
\be\label{Q-LINEAR}
	Q = c_1 p_1 + c_2 p_2 + c_3 \;, \qquad c_j \equiv c_j(x^\LCp,x^\LCm,x^\LCperp,p_\LCm) \;.
\ee
Demanding that the time derivative (\ref{utveckling-lf}) is zero, one obtains an expression cubic in the $p_\LCperp$ which should vanish. Equating coefficients of powers of $p_\LCperp$ gives a series of simple \textit{ordinary} differential equations which specify the coefficient functions $c_j$; one recovers the Poincar\'e generators $Q_1, \ldots, Q_4$ above, as well as the non-Poincar\'e generator
\be\label{Q5}
	Q_5 = \frac{x}{x^\LCp} + Q_1\int\limits^{x^\LCp}\!\ud s\, \frac{1}{s^2 (2 p_\LCm + f(s))} \;,
\ee
which in general is not polynomial in $p_\LCm$. The apparent asymmetry between $x$ and $y$ is due to our choice of basis of independent $Q$'s: instead of $Q_4=L_z$ we could equivalently take
\be\label{Q4tilde}
	\widetilde{Q}_4 =  \frac{y}{x^\LCp} + Q_2\int\limits^{x^\LCp}\!\ud s\, \frac{1}{s^2 (2 p_\LCm + f(s))} \;,
\ee
since
\be
	Q_2 Q_5 - Q_1 \widetilde{Q}_4 = Q_4 \;.
\ee
The explicit solution of the equations of motion proceeds as follows, and is almost entirely algebraic, as for the plane wave case above. It is convenient to use (\ref{Q4tilde}) rather than (\ref{Q4}) as part of the set of independent quantities. From the conservation of (\ref{Q5}) and (\ref{Q4tilde}) we read off the transverse coordinates,
\be\begin{split}
	x(x^\LCp) &= Q_5 x^\LCp - Q_1 x^\LCp \int\limits^{x^\LCp}\!\ud s\, \frac{1}{s^2 (2 p_\LCm + f(s))} \;, \\
	y(x^\LCp) &= \tilde{Q}_4 x^\LCp - Q_2 x^\LCp \int\limits^{x^\LCp}\!\ud s\, \frac{1}{s^2 (2 p_\LCm + f(s))} \;.
\end{split}
\ee
Now that these have been determined, we can read off the transverse momenta from~(\ref{Q123}),
\be
	p_1(x^\LCp) = \frac{Q_1 - 2x(x^\LCp) p_\LCm}{x^\LCp} \;, \qquad p_2(x^\LCp) = \frac{Q_2 - 2y(x^\LCp) p_\LCm}{x^\LCp} \;.
\ee
Since $p_\LCm$ is conserved, it remains only to solve for $x^\LCm$. Hamilton's equations for $x^\LCm$ give
\be\label{decoupled}
	\frac{\ud x^\LCm}{\ud x^\LCp} = -\{x^\LCm,H\} = \frac{1+\big(p_\LCperp(x^\LCp) - A_\LCperp(x^\LCp)\big)^2}{(2p_\LCm +f(x^\LCp))^2} \;.
\ee
Everything on the right hand side has been determined explicitly as a function of $x^\LCp$, so the equation can be integrated directly -- this is the only integral needed, all other coordinates and momenta have been determined algebraically. 

\subsection{The undulator}
%
In this example we consider the spatially oscillating magnetic field
\be \label{UND}
	{\bf B} = B_0(\cos \omega z , \sin\omega z,0) \;,
\ee
modelling a helical undulator. Similarities and differences between relativistic dynamics in undulator and plane wave fields have recently been discussed in \cite{Heinzl:2016kzb}. Non-relativistic charge dynamics in the field (\ref{UND}) was shown to be superintegrable in~\cite{Marchesiello:2015}. One may choose a gauge in which the only non-vanishing components of $A_\mu$ are transverse,
\be
	A_1 = b_0\cos (\omega z) \;, \qquad A_2 =  b_0 \sin (\omega z) \;, \qquad b_0 \equiv \frac{B_0}{\omega} \;.
\ee
In this system, energy $p_0$ is conserved, making it convenient to work in the instant form with Hamiltonian
\be
	H =p_0 = \sqrt{1 + p_3^2 + (p_1 - b_0 \cos(\omega z))^2 + (p_2 - b_0 \sin(\omega z))^2} \;,
\ee
which is time-independent and conserved. As in the non-relativistic limit of this system, the transverse momenta remain conserved,
\be
	\{p_1,H\} =\{p_2,H\} = 0 \;.
\ee
The set $\{H,p_1,p_2\}$ are three independent integrals in involution, and the system is integrable,
\be
	Q_1 = p_1\;, \qquad Q_2 = p_2 \;, \qquad Q_3 = H \;.
\ee
 A fourth Poincar\'e integral, not in involution with $p_1$ and $p_2$, is given by the helical generator
\be
	Q_4 = p_3 + \omega L_z =  p_3 + \omega (x p_2  -y p_1 )\;,
\ee
leading to minimal superintegrability. The four integrals $Q_1\ldots Q_4$ are the relativistic generalisations of those found in the non-relativistic limit~\cite{Marchesiello:2015}. The same is true for a fifth integral, which can be identified as follows. The equations of motion for $y$, $z$ and $p_3$ are, writing a dot for a time ($t$) derivative, 
\be\begin{split}\label{yzp}
	\dot{y} &= -\frac{\partial H}{\partial p_2} = -\frac{p_2- b_0\sin\omega z}{H} \;, \\
	\dot{z} &= -\frac{\partial H}{\partial p_3} = -\frac{p_3}{H} \;, \\
	\dot{p}_3 &=\frac{\partial H}{\partial z_3} = \frac{\omega}{H}\big( p_1 \sin\omega z - p_2 \cos\omega z\big) \;.
\end{split}
\ee
The only difference between these expressions and their nonrelativistic limit resides in the factors of (conserved) $H$ in the denominators. Combining the three equations in (\ref{yzp}) one arrives at
\be\label{yz}
	\frac{\ud y}{p_2 - b_0\sin \omega z} = \frac{\ud z}{p_3} = \frac{\ud z}{\sqrt{2b_0(p_1 \cos\omega z+p_2\sin\omega z) + u}} \;,
\ee
in which $u=Q_3^2-Q_1^2-Q_2^2-1-b_0^2 = const$. This implies, as in the non-relativistic limit~\cite{Marchesiello:2015}, the existence of a fifth conserved quantity which is non-polynomial in the canonical momenta.  Noting that the extra factors of $H$ in (\ref{yzp}) have dropped out, the only difference between (\ref{yz}) and its nonrelativistic limit is in the definition of $u$. Consistency is easily verified -- using that
\be
	Q_3^2 - 1 = H^2  -1  = 2H_\text{non-rel} + \text{relativistic corrections,} 
\ee
recovers, in the non-relativistic limit, the definition of $u$ in~\cite{Marchesiello:2015}.

\section{Superintegrable systems with fewer than three Poincar\'e symmetries}
\label{SECT:SUPER2}
In the examples above superintegrability was realised in terms of Poincar\'e symmetries; here we present two examples in which the number of Poincar\'e symmetries is not enough even to give integrability, but in which there exist additional symmetries in phase space analogous to the Laplace-Runge-Lenz vector of the Kepler problem, see~\cite{Miller:2013,Evans:1990} and references therein. (For a historical account see~\cite{Goldstein:history:1,Goldstein:history:2}.) 
%
%
\subsection{Helical boosts}
%
We return to the front form and consider the electromagnetic fields
\be
\begin{split}
	{\bf E} &= F_0 \big( y , x- \omega x^{\LCp 2} , 0 \big)  \;, \qquad {\bf B} = F_0 \big(  x - \omega x^{\LCp 2}, -y , 0\big) \;,
\end{split}
\ee
where $F_0$ is a constant. The fields are given by a potential with nonzero components
\be\label{A-HEL}
	A_1 = F_0 x^\LCp y \;,  \qquad A_2 = F_0 x^\LCp (x - \frac{\omega}{3}x^{\LCp2}) \;.
\ee
Taking the Lie derivative of the electromagnetic fields with respect to a general Poincar\'e transformation $\xi$, one finds that there are only two Poincar\'e symmetries, $\mathcal{L}_\xi F_{\mu\nu}=0$, for $\xi$ such that $\xi.p = p_\LCm$ or $\xi.p = p_\LCp + 2\omega T_1$, the latter of which might be called a generator of `helical boosts'. The two corresponding conserved quantities are 
\be\label{BOOST:QQ}
	Q_1 = p_\LCm \;, \qquad \widetilde{Q}_2 = H + 2\omega (2 p_\LCm x + x^\LCp p_1) - F_0 y (x + \omega x^{\LCp2}) \;,
\ee
where the final term results from a gauge term in the transformation of (\ref{A-HEL}). The system is, though, superintegrable, which we can show by searching directly for conserved quantities $Q$, making the same ansatz as in (\ref{Q-LINEAR}). Imposing time dependence of $Q$, one finds five conserved quantities with lengthy expressions. To compactify them define
\be
	\Omega = \sqrt{\frac{F_0}{(2p_\LCm)}} \;, \quad \Delta_x = x-y, \quad \Sigma_x = x+y, \quad \Delta_p = \frac{p_1-p_2}{2p_\LCm}, \quad \Sigma_p = \frac{p_1+p_2}{2 p_\LCm} \;.
\ee
The five integrals are then
\be
	\begin{split}
	Q_1 &= p_\LCm \;, \\
	Q_2 &= \Omega  \left(\Sigma_x-\omega{x^{\LCp2}}-\frac{2 \omega }{\Omega^2 }\right)\sinh \Omega x^\LCp + \bigg(\Sigma_p+ {x^\LCp} \left(\Omega ^2 \left(\tfrac{\omega}{3}{x^\LCp}^2 -\Sigma_x\right)+2 \omega \right)\bigg) \cosh \Omega x^\LCp \,,\\
	Q_3 &= \Omega\bigg(  \Sigma_x-\omega{x^{\LCp2}} -\frac{2 \omega }{\Omega^2}\bigg)\cosh \Omega x^\LCp + \bigg(\Sigma_p+ {x^\LCp} \left(\Omega ^2 \left(\tfrac{\omega}{3}{x^\LCp}^2 -\Sigma_x\right)+2 \omega \right)\bigg) \sinh \Omega x^\LCp \,, \\
	Q_4 &= \Omega \bigg( \Delta_x-\omega{x^{\LCp2}} +\frac{2 \omega }{\Omega^2 }\bigg)\cos \Omega x^\LCp + \bigg(\Delta_p- {x^\LCp} (\Omega ^2 \left(\tfrac{\omega}{3}{x^{\LCp2}}  - \Delta_x\right)-2\omega)\bigg)\sin\Omega x^\LCp \,, \\
	Q_5 &= \Omega \bigg(\Delta_x  -\omega{x^{\LCp2}} +\frac{2 \omega }{\Omega^2}\bigg) \sin \Omega x^\LCp - \bigg(\Delta_p- {x^\LCp} (\Omega ^2 \left(\tfrac{\omega}{3}{x^{\LCp2}} -\Delta_x\right)-2\omega)\bigg)\cos\Omega x^\LCp \;,
\end{split}
\ee
depending nonlinearly on $p_\LCm$. The helical boost generator in (\ref{BOOST:QQ}), which is quadratic in $p_\LCperp$, is a combination of these five,
\be
	\widetilde{Q}_2 = \frac{Q_1}{2}(Q_2^2 - Q_3^2 + Q_4^2 + Q_5^2) 
	+ \frac{1}{4Q_1} \; .
\ee
Note that $Q_1\ldots Q_5$ are linear in the transverse momenta, which suggests that the Lorentz boost symmetry $\widetilde{Q}_2$ in (\ref{BOOST:QQ}), quadratic in the transverse momenta, is perhaps less  `fundamental'.
%
\subsection{Vortex beams}
%
We continue to work in the front form. The potential of an electromagnetic vortex has the nonzero components~\cite{BialynickiBirula:2004ev}
\be
	A_1 = B_0 (x\sin\phi-y \cos\phi) \;,	\quad A_2 = B_0 (-x\cos\phi-y\sin\phi) \;,
\ee
where $\phi:=\omega x^\LCp$ ($\omega$ is a frequency scale) and $B_0$ is an amplitude with units of mass squared. The front form Hamiltonian is (\ref{H-LF}) with $A_\LCpm=0$. As in the previous subsection, there are only two Poincar\'e symmetries, under which the potential is invariant, giving the two conserved quantities
\be
	Q_1 = p_\LCm \;, \qquad Q_2 = H + \frac{\omega}{2}(x p_2 - y p_1) \;,
\ee
the second of which is quadratic in the transverse momenta by virtue of the form of the light-front Hamiltonian. Both the classical and quantum equations of motion can be solved exactly in this field~\cite{BialynickiBirula:2004ev}; we will demonstrate here that the classical system is superintegrable.

We again search for conserved quantities $Q$ using the ansatz (\ref{Q-LINEAR}). Imposing the time-independence of $Q$, (\ref{utveckling-lf}), shows that
\begin{enumerate}
	\item $c_1\equiv c_1(x^\LCp,p_\LCm)$,
	\item $c_2$ is determined by $c_1$ (and its time derivatives),
	\item $c_3$ depends linearly on $x$ and $y$, but is otherwise determined by $c_1$, apart from an arbitrary additive function of the conserved $p_\LCm$,
	\item $c_1\equiv c_1(x^\LCp,p_\LCm)$ is itself determined by a {\it fourth order} ordinary differential equation in $x^\LCp$, with $p_\LCm$-dependent coefficients.
\end{enumerate}
Thus there exist four possible independent $c_1$. Along with $p_\LCm$ this already implies a total of five independently conserved quantities, and the system is maximally superintegrable. Rather than give the explicit form of these $Q$, it is interesting to relate the superintegrability of the system to the method by which the equations of motion were solved in~\cite{BialynickiBirula:2004ev}.

Due to the conservation of $p_\LCm$ and the form of the field, the equations of motion for $\{x,y,p_1,p_2\}$ decouple from those for $x^\LCm$. If the `transverse' subsystem of equations for $\{x,y,p_1,p_2\}$ is soluble (for example if it is integrable), then Hamilton's equations for~$x^\LCm$,
\be\label{BB-XM-EOM}
	\frac{\ud x^\LCm}{\ud x^\LCp} = -\{x^\LCm,H\} = \frac{1}{p_\LCm} H \;,
\ee
can be integrated since the Hamiltonian will be determined explicitly in terms of the transverse orbit and the conserved $p_\LCm$, as in (\ref{decoupled}). To pursue this idea, write $z=x+i y$, and a prime for a derivative with respect to $\phi$. The decoupled transverse equations of motion then reduce to~\cite{BialynickiBirula:2004ev,Heinzl:2008rh}
\be
	z'' = \epsilon  e^{i\phi}\bar{z} \;, \qquad \epsilon:= \frac{eB_0}{2\omega p_\LCm} \;,
\ee
where $\epsilon$ will act as an effective coupling. Following~\cite{BialynickiBirula:2004ev} we trade $z$ for a new variable $\chi$ defined by
\be
	z =e^{i\phi/2} \chi \;,
\ee
which makes the system autonomous, as in terms of $\chi$ the equations of motion become
\be
	\chi'' = -i\chi' + \frac{1}{4}\chi + \epsilon {\bar\chi} \;.
\ee
Taking real and imaginary parts of this equation for $\chi=:\alpha+i\beta$, the equations are seen to describe two coupled oscillators, defining $\epsilon_\LCpm  = \frac{1}{4} \pm \epsilon$,
\be
	\alpha'' =\beta ' + \epsilon_\LCp \alpha \;, \qquad \beta'' = -\alpha ' + \epsilon_\LCm \beta \;.
\ee
These equations can be derived from a `nonrelativistic' action $S_E$, where $\{\alpha,\beta\}$ are Cartesian coordinates on a plane and $\phi=\omega x^\LCp$ acts as time,
\be
 	S_E = \frac{1}{2}\int\!\ud\phi\  {\alpha'}^2+{\beta'}^2  +\alpha \beta' - \beta \alpha' +\epsilon_\LCp \alpha^2 + \epsilon_\LCm \beta^2 \;.
\ee
The canonical momenta are
\be\begin{split}
	p_\alpha &= \alpha ' -\frac{1}{2}\beta \;, \qquad p_\beta = \beta ' +\frac{1}{2}\alpha \;,
\end{split}
\ee
and the Hamiltonian is,
\be
	H_E = \frac{1}{2}\bigg[(p_\alpha+\tfrac{1}{2}\beta)^2 + (p_\beta-\tfrac{1}{2}\alpha)^2 - \epsilon_\LCp \alpha^2 - \epsilon_\LCm \beta^2 \bigg]\;.
\ee
The Hamiltonian $H_E$ is conserved\footnote{Our $H_E$ is equivalent to the Hamiltonian in~\cite{BialynickiBirula:2004ev}, but not equal to it. The difference is due to using integration by parts to simplify the action $S_E$, which does not affect the equation of motion but which does lead to a different definition of the canonical momenta.}. We can now search for conserved quantities of this new Hamiltonian system. Let these be $X$, rather than $Q$. One finds here that there are no conserved quantities linear in the momenta $\{p_\alpha,p_\beta\}$, but that there are conserved quantities quadratic in the momenta,
\be\begin{split}
	X_1 &= (p_\alpha+\tfrac{1}{2}\beta)^2 - \epsilon_\LCp \alpha^2 - \frac{\epsilon_\LCp}{\epsilon}\alpha p_\beta + \frac{\epsilon_\LCm}{\epsilon}\beta p_\alpha   \;, \\
	X_2 &= (p_\beta-\tfrac{1}{2}\alpha)^2 - \epsilon_\LCm \beta^2 + \frac{\epsilon_\LCp}{\epsilon}\alpha p_\beta - \frac{\epsilon_\LCm}{\epsilon}\beta p_\alpha   \;.
\end{split}
\ee
These are in involution with the Hamiltonian (by construction) and with each other,
\be
	\{H,X_1\} = \{H,X_2\} = \{X_1,X_2\} = 0 \;.
\ee
Note that the $X_j$ contain terms nonperturbative in the coupling, and that their sum is
\be
	X_1 + X_2 = 2 H_E \;.
\ee
Thus we have two independent conserved quantities in involution, making the transverse system of coupled oscillators integrable. The explicit solutions are given in~\cite{BialynickiBirula:2004ev}, in terms of the original coordinates. With this one solves for the remaining coordinate $x^\LCm$ using~(\ref{BB-XM-EOM}). Finally, it is interesting to note that
\be
	H_E = Q_2 + \text{constant} \;,
\ee
so that the change of variables from $\{x,y\}$ to $\{\alpha,\beta\}$ corresponds to choosing $Q_2$, conserved, as a new Hamiltonian.

\section{Discussion and conclusions}\label{SECT:CONCS}
%
\subsection{Choice of Background}
%
The construction of further superintegrable relativistic systems begins with the choice of background field. In many physical applications it may be desirable for the field, as well as having as many symmetries as possible, to obey Maxwell's equations in vacuum, i.e.~to obey the wave equation $\Box A_\mu = 0$ (in Lorenz gauge, $\partial.A = 0$). An example of such a field is a plane wave, discussed above. To generalise this, consider first the scalar wave equation, $\Box \Phi = 0$, and make the additional assumption that the solution has the particular product form $\Phi(x) = \exp(ik_\LCm x^\LCm) \Psi(x^\LCp, x^{\LCperp})$, with a plane wave phase factor separated off. This implies that the field $\Psi$ obeys
\be \label{SCH}
  (4ik_\LCm \partial_\LCp - \triangle_\LCperp) \Psi = 0 \;,
\ee
which is the 2D Schr\"odinger equation with light-front time $x^\LCp$ as  the time coordinate. The 2D Schr\"odinger equation has been extensively studied by Miller and collaborators as reviewed in~\cite{Miller:1977}, which shows how the symmetry group of (\ref{SCH}), the 2D Schr\"odinger group $Sch(2)$~\cite{Niederer:1972zz}, can be used to classify and determine its solutions. (See~\cite{Son:2008ye,Guica:2010sw,Taylor:2015glc} for recent applications of $Sch(2)$ in the context of holography.)  There are precisely~17 co-ordinate systems for which (\ref{SCH}) separates, the simplest yielding plane wave solutions for $\Psi$. Important for our discussion is the possible impact of the $Sch(2)$ symmetry on the integrability of the charge equation of motion~(\ref{LORENTZ}): $Sch(2)$ is made up of 5 transformations acting on the transverse $(xy)$ plane (two translations, a rotation around the $z$ axis and two null rotations corresponding to Galilei boosts), a time translation in $x^\LCp$, a dilation, a conformal transformation and the identity. The first six of these nine generators form a Galilei subgroup of the Poincar\'e group as has long been known in the context of light-front field quantisation \cite{Susskind:1967rg,Bardakci:1969dv,Kogut:1969xa}. Dilations and conformal transformations are symmetries of massless particles and hence are not shared by the massive charge obeying the Lorentz equation (\ref{LORENTZ}). 

Nevertheless, focussing on Poincar\'e  generators only, it seems straightforward to find a background obeying the wave equation \emph{and} having multiple Poincar\'e symmetries. However, one will typically lose a number of these symmetries upon generalising from scalar to vector solutions, $\Phi \to A_\mu$, which is unsurprising as any choice of vector field singles out a preferred direction. The obvious question is thus whether a sufficient number of Poincar\'e symmetries survives. We therefore plan to  analyse Miller's list of 17 coordinate systems~\cite{Miller:1977} in order to identify backgrounds obeying the wave equation with sufficiently many symmetries to give (super)integrability.

\subsection{Conclusions}
It has long been known in the physics community that relativistic charge motion in a background electromagnetic plane wave is exactly solvable. To the best of our knowledge it has not though been pointed out that this is in fact a maximally superintegrable system.
 
We have presented several further examples of maximally superintegrable systems in background fields. These examples are relativistic, and the backgrounds depend nontrivially on both space and time.
 
It has been conjectured that all maximally superintegrable systems are also exactly solvable quantum mechanically~\cite{Tempesta:2001}. The appropriate quantum extension of our results is not though to quantum mechanics, but quantum field theory, as the quantum theory of a single relativistic particle suffers from e.g.~the Klein paradox. Nevertheless, the first step in such a programme is to solve the Klein-Gordon or Dirac equations in the given background, in order to obtain `first quantised' wavefunctions which provide the input needed for scattering calculations.  In this sense the superintegrable systems we have presented here are indeed also solvable quantum mechanically: the plane wave case is well known~\cite{Volkov:1935,Berestetsky:1982aq}, the vortex beam case has been solved in~\cite{BialynickiBirula:2004ev}, and we solve the TM case in~\cite{Heinzl:2017}.

\ack
We gratefully acknowledge illuminating discussions with Ben King, David McMullan and Marika Taylor. This project has received funding from the European Union's Horizon 2020 research and innovation programme under the Marie Sk\l odowska-Curie grant agreement No.~701676. 

\section*{References}

\end{document}